\documentclass[letterpaper]{article} 
\usepackage[]{aaai27_arxiv}  
\usepackage[hyphens]{url}  
\usepackage{graphicx} 
\def\UrlFont{\rm}  
\usepackage{natbib}  
\usepackage{caption} 
\usepackage{algorithm}
\usepackage{algorithmic}

\newcommand{\magda}[1]{{\color{red}#1}}
\newcommand{\gi}[1]{{\color{teal}#1}}
\newcommand{\yujin}[1]{{\color{violet}#1}}

\usepackage{newfloat}
\usepackage{listings}
\DeclareCaptionStyle{ruled}{labelfont=normalfont,labelsep=colon,strut=off} 
\floatstyle{ruled}
\newfloat{listing}{tb}{lst}{}
\floatname{listing}{Listing}

\usepackage{booktabs}
\usepackage{amsmath}
\usepackage{amssymb}

\usepackage{graphicx}
\usepackage{xcolor}
\usepackage{pifont}

\definecolor{mygreen}{RGB}{0,176,80}
\definecolor{myred}{RGB}{220,38,38}
\definecolor{myorange}{RGB}{230,159,0}

\title{TRACE: Trustworthy Reasoning from Acoustic Cues and Evidence}

\author {
    Giovana Morais\textsuperscript{\rm 1}\corresponding,
    Yujin Kim\textsuperscript{\rm 1},
    Atilay Kucukoglu\textsuperscript{\rm 1,\rm 2},
    Magdalena Fuentes\textsuperscript{\rm 1,\rm 2}
}
\affiliations {
    \textsuperscript{\rm 1}Music and Audio Research Lab, NYU\\
    \textsuperscript{\rm 2}Integrated Design and Media, NYU\\
    giovana.morais@nyu.edu
}

\begin{document}

\maketitle

\begin{abstract}
We introduce \textbf{TRACE} (\emph{Trustworthy Reasoning from Acoustic Cues and Evidence}), a benchmark for diagnosing the reliability of large audio-language models (LALMs) by revealing when, and under what acoustic conditions these models behave untrustworthily. Existing benchmarks primarily measure whether a model produces the correct answer, but correctness alone cannot determine whether the answer is grounded in acoustic evidence or instead reflects language and contextual priors. TRACE evaluates model behavior along three complementary dimensions: (1) \emph{task performance}, whether the answer is correct; (2) \emph{evidence alignment}, whether the model is attending to the audio regions required to answer the question; and (3) \emph{robustness}, whether answers and attributed evidence respond coherently to controlled changes such as event order, overlap, signal-to-noise ratio, or background content. TRACE targets temporal ordering, counting, and event existence, three tasks that remain challenging for current models. We synthesize contextually informed acoustic scenes that allow us to test whether models hallucinate events associated with the context but absent from the audio. This synthesis also allows us to obtain a scalable benchmark for which we know the answers by design. For each scene, we create controlled variants in a programmatic way to stress-test model robustness. We introduce attribution-based metrics that quantify model attention at scale on target events and non-target events. Evaluation across five state-of-the-art models shows that current LALMs often detect where relevant sound evidence occurs, but they do not reliably interpret it, integrate it into the answer, or respond to controlled changes in that evidence. Our benchmark also shows that models still struggle the most with ordering and counting questions.
\end{abstract}


\section{Introduction}

Large Audio Language Models (LALMs) extend Large Language Models (LLMs) by enabling them to reason in natural language about sound, from environmental scenes to
music~\cite{chu2023qwenaudio,ghosh2024gama,liu2023mullama,deng2024musilingo,gong2024ltu,tang2024salmonn, Kong2024AudioFlamingo, kimiteam2025kimiaudio}. They have the potential to unlock a new generation of applications that were previously out of reach: natural language audio search and editing, description of complex auditory scenes, trigger user alerts to relevant auditory events, reasoning about acoustic contexts to drive interactive agents among many others. 

Despite their promise, LALMs perform poorly on recent benchmarks \cite{weck2024muchomusic,sakshi2024mmau, audiotrust2025, ahabench2025, hhjustin, wang2024audiobench, cmibench2025}, struggling even with simple temporal reasoning tasks, such as understanding the order of sound events (e.g. sound $A$ happened before sound $B$)~\cite{hhjustin}, or counting sound events in the audio~\cite{sakshi2024mmau}, though it is unclear \textit{why} \cite{morais2026ashap}. One possible explanation is that Audio LLMs may not consistently use the audio signal. In several cases, replacing the input audio with random noise does not significantly change model performance~\cite{weck2024muchomusic, sakshi2024mmau}. Furthermore, LALMs have been shown to exhibit significant acoustic hallucinations when they are unsure~\cite{ahabench2025, audiotrust2025}. This is an issue for both performance and trustworthiness. For many applications deploying models which can hallucinate sounds and do not reliably listen to the audio to answer questions is a major risk. 

Existing audio-language benchmarks primarily evaluate \emph{correctness}: whether a final answer matches a reference \cite{sakshi2024mmau, weck2024muchomusic, wang2024audiobench, cmibench2025, hhjustin, audiotrust2025}. Correctness is necessary, but it does not establish that the model relied on acoustic evidence.
A correct answer may instead reflect other sounds associated with the scene or priors learned from language. We have limited insight into why a model fails, or whether its answers are robust to irrelevant acoustic degradation. Recent audio benchmarks have begun to measure hallucination, safety, and broad task performance at scale \cite{audiotrust2025, ahabench2025}, but they largely evaluate answer correctness rather than its evidentiary basis \cite{wang2024audiobench, cmibench2025, kuan2024hallucinationsmultitask,
zang2025reallylisteningboostingperceptual}.

Unlocking the next level of AI-powered interaction with audio requires models that are \textit{grounded} in acoustic evidence, and \textit{robust} to irrelevant acoustic variation such as channel noise and reverberation.


We introduce \textbf{TRACE}, a diagnostic benchmark for acoustic grounding and robustness in binary audio question answering. TRACE combines controlled soundscape synthesis, programmatic answers, event-level temporal annotations, and novel metrics to quantify these aspects of models' trustworthiness. The benchmark targets existence, ordering, and counting because these tasks require localized acoustic evidence and remain difficult for current systems. Scene composition is sampled from foreground-foreground and foreground-background co-occurrence statistics estimated from AudioSet Strong eval~\cite{hershey2021audiosetstrong,audioset}. This produces contextually informed mixtures and supports negative existence questions about events that are plausible in the scene but absent from the signal.


Our contributions are: 1) A benchmark of contextually informed multi-event scenes with programmatic questions for existence, ordering, and counting; 2) Grounding metrics that quantify attribution to target events, distractor events, and regions outside annotated events; 3) Paired robustness tests that distinguish invariance to answer-preserving transformations from responsiveness to evidence-changing interventions; 4) A diagnostic evaluation showing how models with different answer accuracy can have similar grounding in acoustic events, and that most errors occur despite the correct evidence localization.

\section{Related Work}

\paragraph{Audio-language benchmarks.}
AudioBench, MMAU, MuChoMusic, AudioTrust, and AHa-Bench evaluate broad audio understanding, reasoning, hallucination, or trustworthiness~\cite{wang2024audiobench,sakshi2024mmau,weck2024muchomusic,audiotrust2025,ahabench2025}. These benchmarks reveal low performance especially on temporal and relational tasks such as ordering, counting, and overlap~\cite{hhjustin}, but provide limited insight into why models fail. Importantly, they show that  performance remains similar for some models when audio is replaced by noise or when compared with text-only systems, suggesting that models rely on language priors over acoustic evidence~\cite{reid2024geminipro,zang2025reallylisteningboostingperceptual}. Furthermore, models have been reported to confidently hallucinate \cite{ahabench2025, audiotrust2025}.

To date, there is no evaluation benchmark that systematically investigates these model pitfalls beyond accuracy alone. This is paramount because there are multiple reasons why models could be failing, and right now they are conflated because better evaluation tools are lacking. TRACE evaluates whether an answer is associated with relevant acoustic evidence and whether that association behaves correctly under acoustic variation.

\paragraph{Explaining audio and multimodal models.}
In computer vision and language, limited insight into model behavior has motivated extensive work on explainability beyond performance metrics \cite{Selvaraju2017GradCAM, parcalabescu2023mmshap, Vig2019BertViz, Sundararajan2017IntegratedGradients, ribeiro2016lime,
lundberg2017shap, hessel2020doesmymodel, lyu2022dime, parcalabescu2024measuring, gat2021perceptual}. In audio, however, explainability remains underexplored \cite{audiolime, lime_dixon, bolanos2025benchmarking, explainable_audio, audio_mnist, fucci2024explainability}, particularly for Audio LLMs \cite{morais2026ashap,sotirou2024musiclime}.

Most explainability methods applied to audio are post-hoc: they mask portions of the input and measure how the prediction changes to produce an importance map \cite{audiolime, bolanos2025benchmarking, lime_dixon}. This provides insights on model behavior by indicating whether the model listened to the audio, and where \cite{morais2026ashap}. Although broadly applicable, these maps are typically inspected manually and lack quantitative metrics for summarizing model behavior \cite{explainable_audio}, limiting their scalability and adoption. We address this gap by measuring and quantifying alignment between model explanation attributions and the event’s temporal ground-truth. Our approach builds on and extends prior work that used Shapley values to quantify LALM reliance on audio and text at the clip level \cite{shapley1953, morais2026ashap}, and temporal annotations to benchmark post-hoc audio explainers \cite{bolanos2025benchmarking}.

\begin{figure*}[t]
    \centering
    \includegraphics[width=\linewidth]{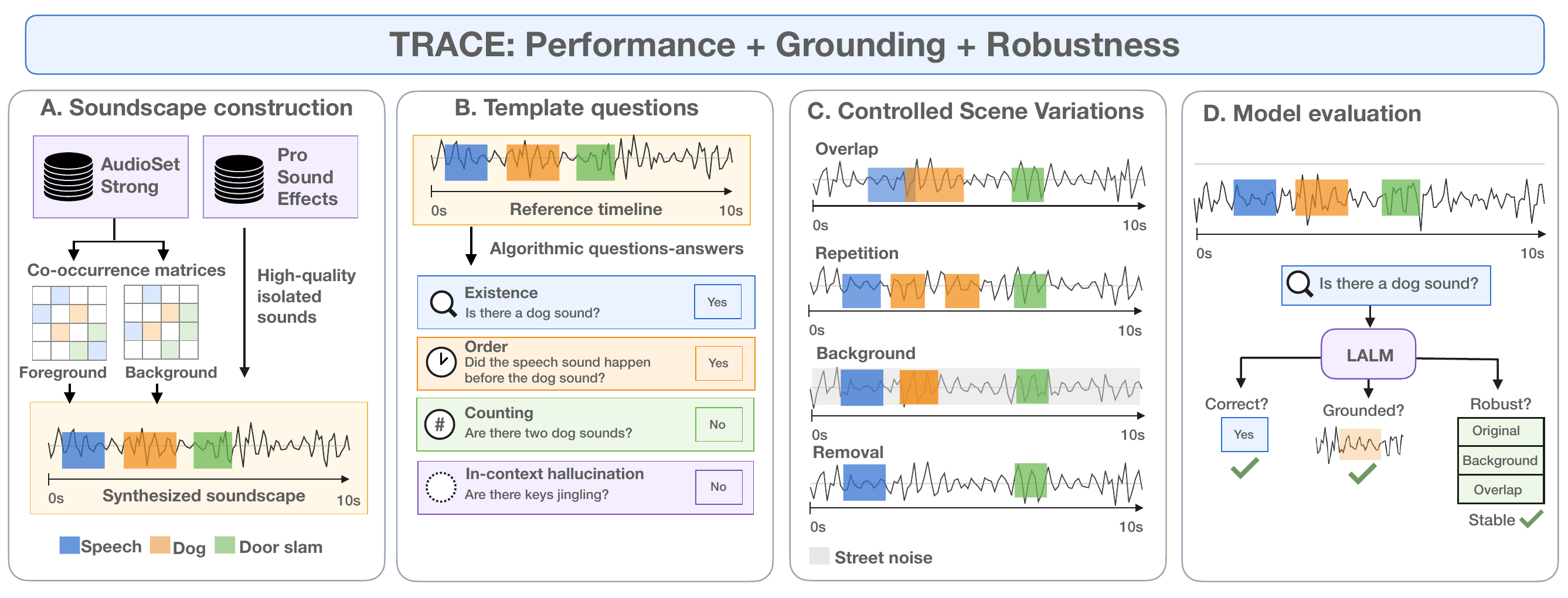}
    \caption{TRACE pipeline. Co-occurrence statistics from temporally annotated recordings guide the composition of multi-event soundscapes from isolated sources. Programmatic templates generate binary existence, ordering, and counting questions together with task-specific temporal evidence. Each reference scene is paired with answer-preserving nuisance transformations and evidence-changing counterfactual interventions. LALMs are evaluated for answer accuracy, attribution to target and distractor regions, and whether their answers and evidence change as expected across transformations.}
    \label{fig:trace_overview}
\end{figure*}

\section{Measuring Grounding and Robustness}

Following \citet{morais2026ashap}, we obtain model explanations using Shapley Values by masking the audio waveform in small windows, and measuring the change in the logits of the original response the model gave without masking. For example, for  a given question "Did the dog sound happened before the door slam sound?", if the model gave an original answer of "Yes", we mask a segment of the audio with zeros (i.e. the small window), and aggregate the change in the output logits of "Yes", "yes", and other affirmative related tokens to determine the model attribution for that masked window. We repeat this process until the audio is fully masked and then start re-adding windows back. This is known as the Permutation approximation and it is implemented in the \texttt{shap} Python library as one option to avoid the full calculation of Shapley values that becomes computationally intractable as the number of audio windows increases.
We aggregate the logits of accepted affirmative and negative tokenizations into a
binary margin $m = L_{\mathrm{yes}} - L_{\mathrm{no}}$ and explain $m$ when the model
answers yes and $-m$ when it answers no, so that positive attribution always supports
the model's selected answer. Attribution sign is therefore anchored to the model's
own decision, not to the correct answer; this is what allows us to identify cases
where a model reads the correct evidence with the wrong polarity.


To assess whether a model bases its answer on relevant acoustic evidence, we compare its Shapley-value explanation with the ground-truth intervals of the target event, which is queried by the question, and the distractor events, which are the other sources in the audio. We summarize this alignment using two quantities: Attribution Mass ($A$) and Attribution Density ($D$) defined below.

\paragraph{Attribution Mass}

Attribution mass measures what fraction of the model’s total attribution falls within a region, such as the target event.
 For a given sample let $G_T$ denote the union of target-event intervals and $G_D$ the union of distractor-event intervals. Let $\phi_t$ be the Shapley value at time $t$, and let $c \in \{+, -, \mathrm{abs}\}$ indicate whether we consider positive, negative, or absolute attribution of the region $e$:
$$
e_t^{(+)}=\max(\phi_t,0), \quad
e_t^{(-)}=\max(-\phi_t,0), \quad
e_t^{(\mathrm{abs})}=|\phi_t|.
$$
We define the corresponding attribution-mass scores as
\begin{equation}
A_T^{(c)}=
\frac{\sum_{t\in G_T} e_t^{(c)}}{\sum_t e_t^{(c)}},
\qquad
A_D^{(c)}=
\frac{\sum_{t\in G_D} e_t^{(c)}}{\sum_t e_t^{(c)}},
\end{equation}
and
\begin{equation}
A_O^{(c)}=\frac{\sum_{t \notin G_T \cup G_D} e_t^{(c)}}{\sum_t e_t^{(c)}},
\qquad
A_\cap^{(c)}=\frac{\sum_{t \in G_T \cap G_D} e_t^{(c)}}{\sum_t e_t^{(c)}}.
\end{equation}
Because target and distractor spans may share analysis windows, $A_T$, $A_D$ and
$A_O$ do not partition the attribution; instead
$A_T^{(c)} + A_D^{(c)} - A_\cap^{(c)} + A_O^{(c)} = 1$.

These quantities measure the fraction of positive, negative, or absolute attribution assigned to target events ($A_T$), distractor events ($A_D$), and regions outside annotated events ($A_O$). We report $A_\cap$ in the supplement. Under uniform attribution, each time point receives the same share of the total mass. Therefore, a target occupying $|G_T|/T$ of the clip receives an expected attribution mass of $\mathbb{E}[A_T^{\mathrm{rand}}]=|G_T|/T$. Values above this baseline indicate that attribution is more concentrated on the target than expected from its duration alone.

\paragraph{Attribution Density Ratio}

Besides the mass, we also report mean attribution density ratio for target and distractor events. Density ratio adjusts attribution mass for region duration, measuring whether attribution is more concentrated in that region than expected from its length. We formally define it as:

\begin{equation}
    D_T^{(c)} = \frac{\frac{1}{|G_T|} \sum_{t\in G_T} e_t^{(c)}}{\frac{1}{T}\sum_{t = 1}^{T} e_t^{(c)}}, 
    \qquad  
    D_D^{(c)} = \frac{\frac{1}{|G_{D}|} \sum_{t\in G_{D}} e_t^{(c)}}{\frac{1}{T}\sum_{t = 1}^{T} e_t^{(c)}}
\end{equation}


$D_T=1$ indicates uniform attribution density (random), with $D_T>1$ and $D_T<1$ denoting concentration above and below random, respectively. Region extents are measured in analysis windows; a window is assigned to a region if
it overlaps that region by at least half its duration.






A model may exhibit high $A_T$ and $D_T$, indicating that a large share of attribution is concentrated on target events. High $A_D$, $D_D$ indicate reliance on distractor events, while high $A_O$ indicates attribution to regions outside annotated events. 
We call a target event grounded when $D_T > 1$, i.e. when
the target region receives more attribution per unit time than the clip
average. Since $D_T = A_T \cdot T / |G_T|$, this is equivalent to the target
attribution mass exceeding the duration baseline $|G_T|/T$.

\subsection{Robustness}

Besides being grounded, a robust model should preserve its evidence allocation under acoustic variations that do not change the correct answer, such as changes in target SNR, reverberation, polyphony, or background noise. We evaluate this using paired reference and perturbed inputs and compute $\Delta A_T = A_T^{\mathrm{pert}} - A_T^{\mathrm{ref}}$, $\Delta A_D = A_D^{\mathrm{pert}} - A_D^{\mathrm{ref}}$, and $\Delta A_O = A_O^{\mathrm{pert}} - A_O^{\mathrm{ref}}$. Attribution mass is suitable for this comparison because it measures how the model distributes its total attribution among target, distractor, and outside-event regions. Since it is normalized by the total attribution, it captures changes in evidence allocation rather than changes in the overall magnitude of the Shapley values. For answer-preserving perturbations, a robust model should maintain similar attribution mass across regions, yielding values close to zero.





\section{TRACE Benchmark}

TRACE exploits controlled environmental-sound synthesis to evaluate accuracy, grounding and robustness, within a unified framework. Its design is guided by three requirements. First, questions and answers must be generated programmatically so that correctness can be established without relying on an additional model or human judge. Second, semantic context must be controllable: the composition of a scene should
follow the co-occurrence structure of real recordings, while the acoustic background
remains neutral by default, so that contextual pressure can be introduced as in a controlled fashion.  Third, the benchmark must provide precise event-level ground truth for evaluating grounding and for generating controlled robustness variations. 

\subsection{Co-occurrence-Informed Soundscape Construction}

To obtain full control over the semantics and relations among environmental-sound
events---their identity, timing, repetition, and overlap---we synthesize multi-event
audio mixtures rather than curating existing recordings. This design yields
event-level annotations that support our grounding and robustness metrics, whereas
most existing audio-reasoning benchmarks provide only clip-level
labels~\cite{weck2024muchomusic,sakshi2024mmau}. Unconstrained combinations of
isolated sounds, however, would produce implausible scenes that do not reflect the
contextual dependencies of real audio. We therefore condition scene composition on
co-occurrence statistics estimated from natural recordings.

We derive these statistics from AudioSet Strong~\cite{hershey2021audiosetstrong},
which provides manual temporal annotations for approximately 17{,}000 user-uploaded
YouTube videos. Because these recordings capture naturally occurring acoustic scenes,
their co-occurrence statistics provide evidence about which events and backgrounds
plausibly appear together. From the annotated class pairs we estimate a
foreground--foreground and a foreground--background co-occurrence matrix. Each class
pair contributes at most once per recording, independent of repetition count, so that
long recordings with repeated events do not dominate the statistics. Row normalization
converts the resulting counts into conditional sampling distributions: the
foreground--foreground distribution guides the selection of event classes within a
scene, and the foreground--background distribution guides the selection of background
conditions. The latter is used when generating background variations as our reference scenes use brown noise as background. 
This keeps the background semantically
neutral in the reference condition, so that recorded ambiences sampled from the
foreground--background distribution can be introduced as a controlled variation.

We draw source material from the Pro Sound Effects library~\cite{prosoundeffects2026},
which provides over 1.2M high-quality, manually labeled isolated recordings whose
taxonomy overlaps substantially with AudioSet~\cite{audioset}. We manually align its onology with AudioSet ontology, mapping fine-grained classes onto their nearest AudioSet ancestor and discarding the classes with no counterpart; which yielded 74 source classes. We aggregate the AudioSet Strong
co-occurrence counts over the same mapping, so that the sampling distributions are
defined on this 74-class vocabulary. We then partition the
recordings into foreground events ($\leq 3$\,s) and background conditions
($\geq 9$\,s) using annotation duration and ontology information, discarding
recordings of intermediate duration; because the library serves only as synthesis
material, this costs us no coverage. Both thresholds are set relative to our 10\,s
scenes, so that up to three foreground events fit without overlap and a single
background covers the majority of the track.

Scene generation happens in three steps: we sample a starter foreground class,
sample additional foreground classes conditioned on it and position the foreground classes on the reference scene, guaranteeing that they don't overlap so we have control of it as a variation type. Source files, and onset times are drawn subject to these class assignments and recorded before
rendering; these records constitute the event-level annotations released with TRACE.
All synthesized scenes are 10\,s, 44.1\,KHz, mono, 32 bit. 
TRACE comprises 1{,}000 reference scenes, each paired with four binary questions,
and 3{,}000 transformed scenes carrying one question each, for a total of 7{,}000
(audio, question) items. 

\subsection{Questions and Task-Relevant Evidence}

In TRACE, each reference scene synthesized is paired with 4 binary questions, resulting in 4000 audio, question pairs. The questions cover three family tasks:

\paragraph{Existence.}
Questions ask whether an event class occurs in the scene. 
Positive questions sample a present class. Negative questions sample an absent class with high co-occurrence with the present events, creating a contextually plausible distractor. For a positive existence question, the target evidence $G_T$ is the union of all intervals containing the queried class. A negative existence question has no target interval; it is evaluated through answer accuracy and distractor attribution.

\paragraph{Ordering.}
Questions ask whether event $a$ occurs before event $b$, whether an event occurs first or last, or whether two events overlap. The target evidence contains the event instances whose temporal relation determines the answer. When multiple instances could satisfy a template, generation rules select a random pair and store its intervals.

\paragraph{Counting.}
Questions ask whether an event occurs exactly, more than, or fewer than a specified number of times. The target evidence is the union of all instances of the queried class because the answer depends on the complete count.

Template variables are instantiated from the annotated timeline, and answers are recomputed after every transformation. This construction provides the answer, the relevant temporal evidence, and the set of non-target event intervals without human or model-based labeling.

\subsection{Controlled Transformations}
\label{ssec:variations}

Each reference scene is paired with task-conditioned transformations. We distinguish transformations by their expected effect.

\paragraph{Answer-preserving nuisance transformations.}
These transformations alter acoustic conditions while preserving the event structure required by the question. They include background substitution or addition, moderate source-to-background ratio changes, gain changes, and temporal overlap changes that do not alter the queried relation. They test whether the answer and attributed evidence remain stable when the relevant event remains present and the ground-truth answer is unchanged.

\paragraph{Evidence-changing counterfactual interventions.}
These transformations modify the event structure and may change the correct answer. They include target insertion or removal for existence, repetition changes for counting, temporal displacement or order reversal for ordering, and overlap changes that alter two events' relation. The transformed answer and target intervals are recomputed from the modified timeline.

This separation is necessary because robustness requires two behaviors: invariance to irrelevant changes and responsiveness to relevant changes. A model that never changes its answer is stable but not robust.

\subsection{Benchmark Statistics}

\begin{table}[t]
\centering
\small
\setlength{\tabcolsep}{4pt}
\begin{tabular}{lrrrrr}
\toprule
& \multicolumn{3}{c}{\textbf{Scenes}}
& \multicolumn{2}{c}{\textbf{Questions}} \\
\cmidrule(lr){2-4}
\cmidrule(l){5-6}
\textbf{Task} & \textbf{Ref.} & \textbf{Nuis.} & \textbf{CF} & \textbf{Yes} & \textbf{No} \\
\midrule
Existence & 2000 & 743 & 257 & 1743 & 1257 \\
Ordering  & 1000 & 766 & 234 & 635 & 1365  \\
Counting  & 1000 & 793 & 207 & 677 & 1323  \\
\midrule
Total     & 4000 & 2302 & 698 & 3055 & 3945 \\
\bottomrule
\end{tabular}
\caption{TRACE statistics. Ref. denotes reference scene; Nuis. denotes answer-preserving nuisance transformations; CF denotes evidence-changing counterfactual interventions. We report the number of scenes by type and distribution of question answers for each task.}
\label{tab:dataset_stats}
\end{table}

Table~\ref{tab:dataset_stats} shows the statistics of the TRACE benchmark. Our benchmark consists of 7,000 scenes, including 4,000 reference scenes and 3,000 transformed scenes. Along with the scenes, we generated 7,000 questions including 1,632 answers as "yes" and 2,368 as "no" for reference scenes, and 1,423 answers as "yes" and 1,577 as "no" for the transformed scenes. The transformed scenes include 797 background transformations, 762 signal-to-noise variations, 511 temporal overlap, 257 event removal, 234 inverse order, 232 distance variations, and 207 repetition transformations. Overall, the benchmark covers 74 sound event classes, rendered from 2,581 unique audio sources from the Pro Sound Effects Library \cite{prosoundeffects2026}, and contains 12,768 event instances across all reference and transformed scenes. Each task (Existence-Positive, Existence-Negative, Ordering, and Counting) contains 1,000 reference scenes, with an additional 1,000 transformation scenes per task except Existence-Negative. Since our variations target events in the audio, applying it to a non-existent event does not make sense and does not bring meaningful information to our analysis.

\section{Experimental Setup}

\subsection{Models and Prompts}

We evaluate Qwen2.5-Omni-7B (Qwen2.5), Qwen3-Omni-30B-A3B-Instruct (Qwen3), Audio Flamingo 3 (AF3), Audio Flamingo Next (AF-Next) and Kimi-Audio-7B-Instruct (Kimi-Audio). Models are selected based on support for non-speech audio, public inference access, and competitive performance on recent audio-language benchmarks. All models receive the same binary question format and a fixed instruction to answer yes or no. We use deterministic decoding for answer evaluation. Model-specific response token sets are defined before evaluation and held fixed across conditions.

\subsection{Attribution Configuration}

The primary explainer is temporal Shapley-value attribution over waveform windows. For this study, we ran a small benchmark to evaluate how feasible the computation was for different audio window sizes, comparing \{10, 20, and 50\} windows for a 10 s audio, also varying the number of permutations we could run over them. We chose 20 audio windows (each 500ms) and 200 evals, meaning that we have 5 permutations over the features. Computational time varied by model, AF-Next being the fastest and Qwen3-Omni the slowest. Our primary results use mean attribution and positive attribution to the selected-answer margin. Negative attribution is retained for supplementary analysis.



\section{Results and Discussion}



\begin{figure}[ht]
    \centering
    \includegraphics[width=\linewidth]{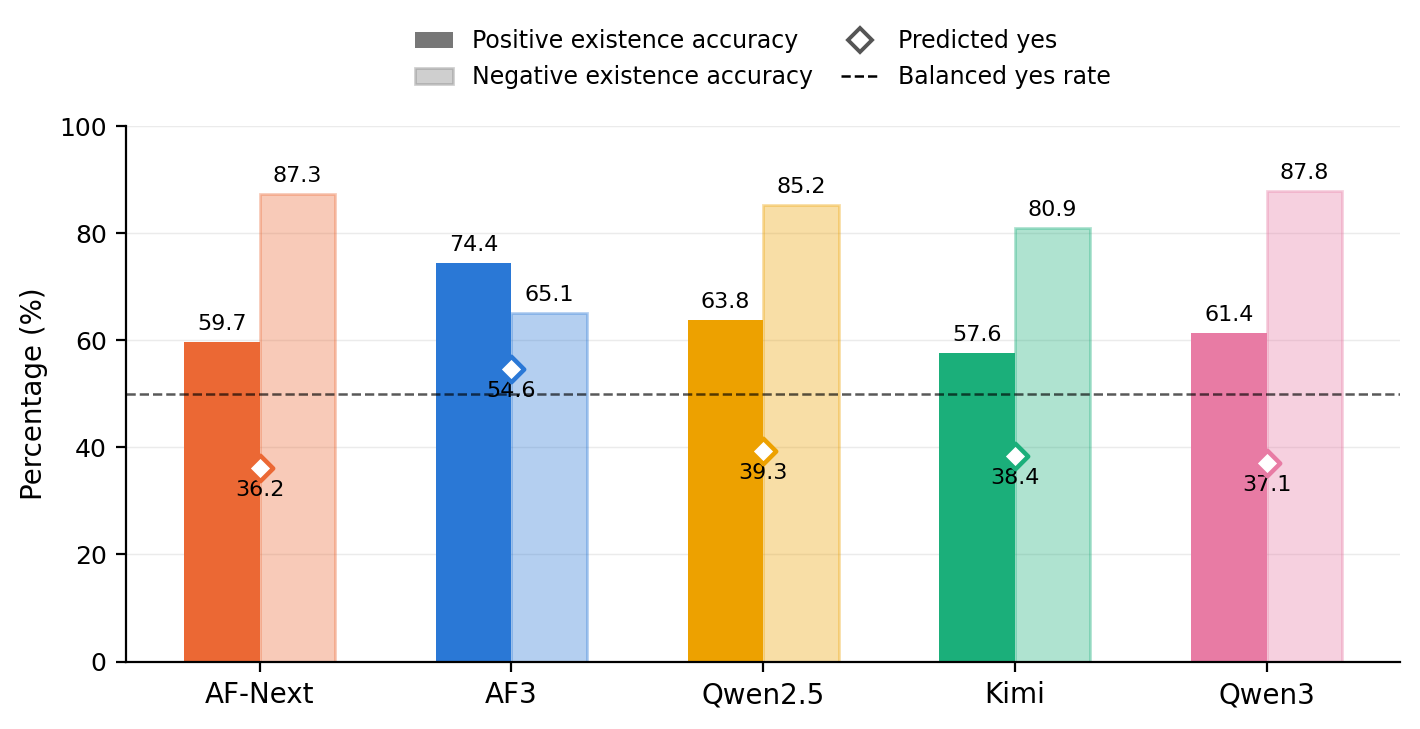}
    \caption{Positive- and negative-existence accuracy by model. Diamonds show the fraction of predictions answered “yes”; the dashed line marks the balanced rate of 50\%. Most models favor negative responses.}
\label{fig:positive_vs_negative}
\end{figure}

Table~\ref{tab:metrics_reference} shows model performance in the three tasks (ordering, counting, and existence), along with the grounding metrics for reference scenes only. Qwen3-Omni exhibits the best performance overall, followed by Kimi, and AF-Next has the lowest performance. Breaking down by question type, this difference is biggest in counting questions, where Qwen3-Omni correctly answers 91\% of them, while the second best model (Kimi) achieves ~65\% accuracy. Order questions are particularly low for AF-Next, which achieves an accuracy of 31\%. Overall, models perform best in questions asking about the existence of objects that are not present in the audio (negative existence questions, with an accuracy of 81\%), than in questions about objects that \textit{are} present in the audio (64.3\% accuracy). This is an interesting and counterintuitive result, given that the negative questions are asking about a sound event that is often heard in the context of the other sounds and this should confuse the models. When looking into more detail, as shown in Figure \ref{fig:positive_vs_negative}, models that perform better in negative existence questions are those that tend to favor negative responses, and thus perform worse in positive existence questions.

\magda{
\begin{table}[t]
\centering
\small
\setlength{\tabcolsep}{4.5pt}
\begin{tabular}{@{}lrrrrr@{}}
\toprule
& \multicolumn{4}{c}{\textbf{Accuracy (\%)}} \\
\cmidrule(lr){2-5}
Model
& Overall
& Order
& Exist.
& Count. \\

\midrule
AF-Next
& 55.0
& 31.2
& 73.5
& 42.0 \\

AF3
& 63.0
& 52.1
& 69.8
& 60.3 \\

Qwen2.5
& 67.8
& 63.5
& \textbf{74.5}
& 58.5 \\

Kimi
& \underline{69.3}
& \underline{70.9}
& 69.3
& \underline{67.9} \\

Qwen3
& \textbf{78.4}
& \textbf{73.3}
& \underline{73.6}
& \textbf{92.9} \\

\midrule
All
& 66.8\%
& 58.2\%
& 72.3\%
& 64.3\% \\
\midrule
Random & 50.0 & 50.0 & 50.0 & 50.0  \\
\bottomrule
\end{tabular}
\caption{
Performance on reference scenes without transformations.
Best model results are in bold and second-best model results are underlined.
The All row pools predictions and attribution records across models.
}
\label{tab:metrics_reference}
\end{table}
}

We also observed that model competence depends on the task. Some examples of this are that Qwen3 reaches 92.9\% accuracy on reference counting but 61.4\% on positive existence, Kimi reaches 70.9\% on ordering but 57.6\% on positive existence, and AF-Next reaches 87.3\% on negative existence but 31.2\% on ordering. Overall, existence questions are the easiest ones for the models, followed by counting and ordering questions. This is on par with the literature, which has recently demonstrated that models have a hard time understanding temporal relationships between acoustic events \cite{bhattacharya25b_interspeech}.




\subsection{Grounding Does Not Explain Cross-Model Accuracy Differences}
Target attribution mass is nearly constant across models (0.259--0.306) while
accuracy spans more than twenty points (55.0--78.4\%). Grounding, as measured by
attribution mass, therefore does not explain the accuracy differences between these
systems: all five allocate a similar share of their attribution to the target region
and differ in what they do with it. 

\begin{figure}[ht]
    \centering
    \includegraphics[width=\linewidth]{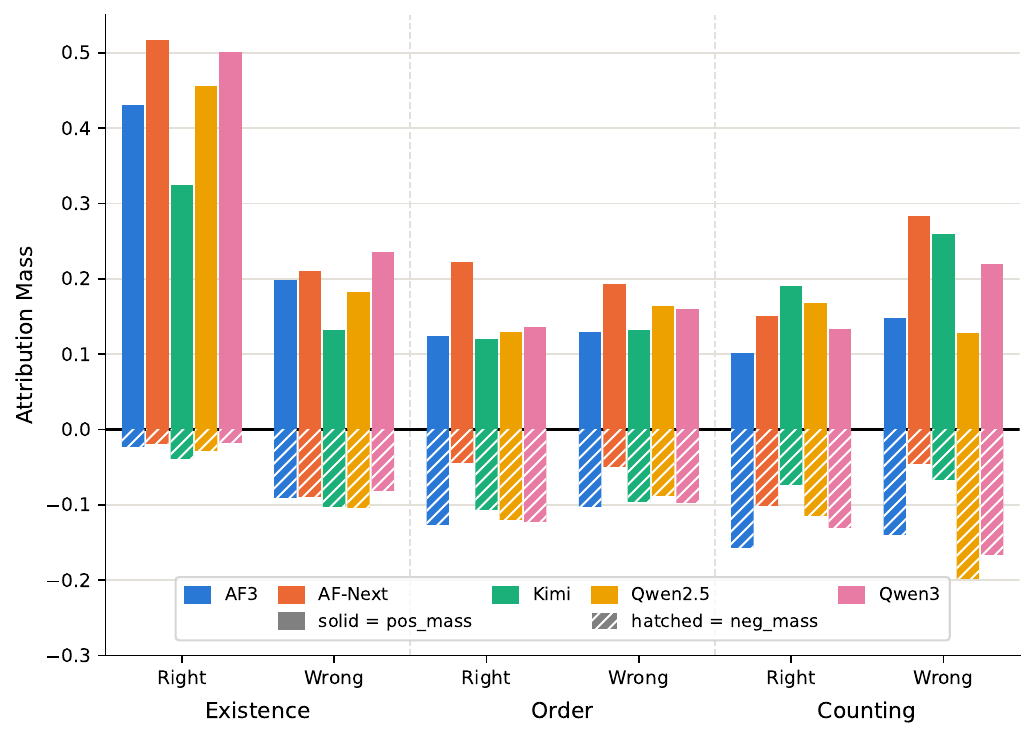}
    \caption{Positive and negative attribution mass on target event. Hatched plot shows the negative mass value. Existence questions show the highest positive attribution mass across models.}
\label{fig:attribution_mass_target}
\end{figure}

All models assign significantly more attribution to target regions than the duration
baseline (Figure~\ref{fig:attribution_mass_target}).




\subsection{Models' Robustness}

\begin{figure}[ht]
    \centering
    \includegraphics[width=\linewidth]{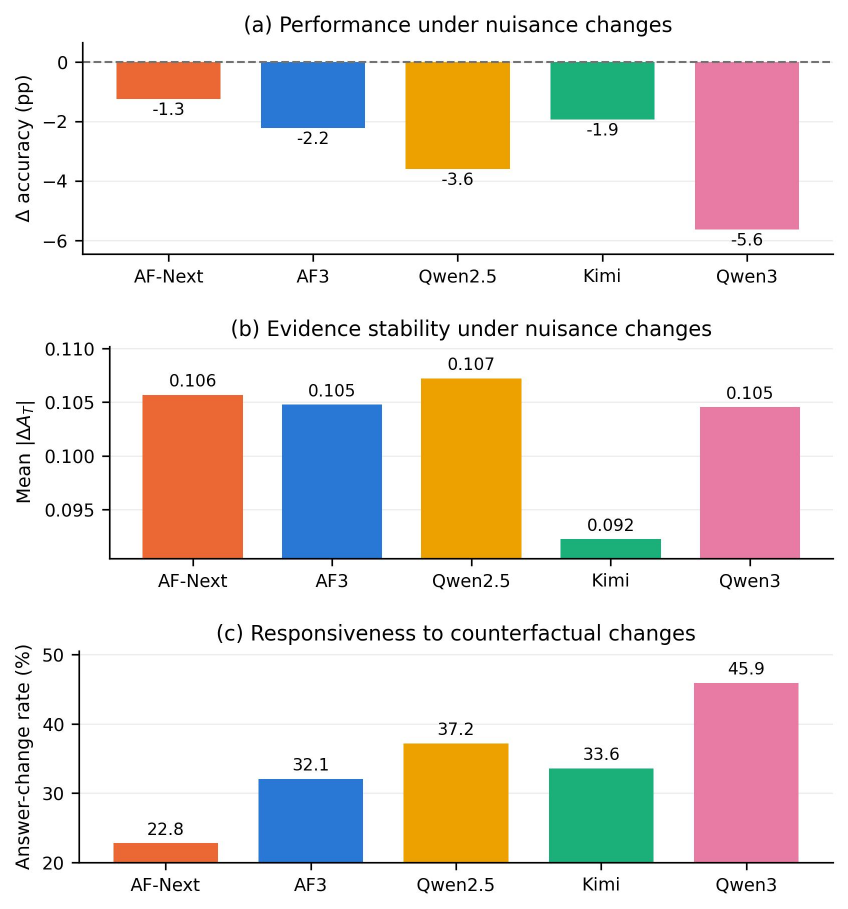}
    \caption{Robustness analysis. How nuisance and counterfactual changes affect model accuracy, response and grounding metrics.}
\label{fig:robustness}
\end{figure}

Figure~\ref{fig:robustness} shows changes in performance, attribution mass and rate in which the models change responses adequately for counterfactual questions. All models lose accuracy under answer-preserving acoustic changes, indicating limited invariance to nuisance variation. AF-Next shows the smallest decrease in accuracy ($-1.25$ percentage points), followed by Kimi ($-1.94$) and AF3 ($-2.22$). Qwen3 shows the largest decrease ($-5.62$), despite achieving the highest accuracy on the reference data.

Changes in target attribution mass are similar across models, with mean $|\Delta A_T|$
between $0.092$ and $0.107$, a range we do not interpret as separating them. Evidence
allocation shifts by a similar, small amount for every system under answer-preserving
change.

For counterfactual changes, Qwen3 is the most responsive, changing its answer in $45.9\%$ of cases where the correct answer changes. AF-Next is the least responsive at $22.8\%$. No model responds correctly to more than half of the counterfactual changes. These results reveal a trade-off between invariance and responsiveness: AF-Next is stable under answer-preserving changes but often fails to react when the answer should change, whereas Qwen3 is more responsive to counterfactual evidence but less stable under nuisance variation.



\subsection{Common Failure Modes}

We investigate some common failure cases we observed empirically and how they are quantified with our metrics. 

\begin{table}[t]
\centering
\small
\setlength{\tabcolsep}{4pt}
\begin{tabular}{lrrrr}
\toprule
& \multicolumn{3}{c}{\textbf{All questions}} & \textbf{Incorrect} \\
\cmidrule(lr){2-4}\cmidrule(l){5-5}
\textbf{Model} & \textbf{Cons.} & \textbf{Contra.} & \textbf{Ungr.} & \textbf{\% grounded} \\
\midrule
AF3      & 0.452 & 0.265 & 0.284 & 66.6 \\
AF-Next  & 0.345 & 0.386 & 0.269 & 71.2 \\
Kimi     & 0.394 & 0.297 & 0.309 & 67.2 \\
Qwen2.5  & 0.414 & 0.317 & 0.269 & 72.2 \\
Qwen3    & 0.488 & 0.236 & 0.276 & 72.5 \\
\bottomrule
\end{tabular}
\caption{Grounded cases split by whether the dominant attribution
sign argues for the correct answer (Consistent) or against it (Contradictory); the three
shares sum to 1. The final column is the fraction of \emph{incorrect} answers
that were still grounded on the target event. Sign is anchored to the model's own answer, so a case is consistent (Cons.) when the
dominant attribution sign supports the answer the model gave and that answer is
correct, and contradictory when the dominant sign supports the model's answer
but the answer is wrong.}
\label{tab:grounding_decomp}
\end{table}

\paragraph{Evidence misinterpretation dominates.}
Across all five models, between 66.6\% and 72.5\% of \emph{incorrect} answers
were nonetheless grounded on the target event (Table~\ref{tab:grounding_decomp}).
Models strongly attend to the acoustic evidence the question depends on and
then fail to convert it into the right answer. In 23.6--38.6\% of all cases the
model is grounded but the dominant sign of its attribution argues for the wrong
answer. 
For AF-Next and Qwen3 target
attribution mass is in fact higher on incorrect answers than correct
ones (0.367 vs.\ 0.307 and 0.364 vs.\ 0.324), suggesting that additional
attention to the right region does not translate into interpretation.

\paragraph{Target neglect is the minority mode.}
Only 26.9--30.9\% of questions are ungrounded. At the population level all models place significantly more mass on target regions
than their duration predicts (mean excess $+0.094$ to $+0.152$; two-sided Wilcoxon
signed-rank $p < 10^{-100}$ for every model and question type). This suggests that event detection is
not the bottleneck.


\paragraph{Models do not over-rely on background.}
Contrary to the expectation that ungrounded answers are driven by background
context, attribution to regions outside annotated events falls \emph{below} the
background's duration share for every model and every acoustic condition (for
example, 0.197--0.270 against a duration share of 0.554 under overlap).
Background share also fails to separate correct from incorrect answers in any
consistent direction. Kimi is the most background-weighted model throughout,
and correspondingly shows the largest ungrounded fraction (30.9\%).

\paragraph{Decision instability.}
Models change their answer on 9.3--20.1\% of variations that leave the correct
answer unchanged, while target attribution mass moves by only
$|\Delta A_T| \approx 0.105$ on average. The answer is therefore unstable under
perturbations that barely move the evidence allocation, indicating that the
decision is not tightly coupled to the attribution pattern that supports it.


\section{Limitations}

TRACE uses synthetic mixtures to obtain exact event structure and paired
interventions. Co-occurrence-informed sampling constrains which events and backgrounds
appear together but does not govern the acoustic realization of a rendered mixture, so
we describe TRACE scenes as \emph{contextually informed} rather than ecologically
valid. Post-hoc attribution is an operational diagnostic rather than a causal account
of model reasoning, and results may depend on the explainer, masking baseline,
temporal resolution, and answer-token construction; logit access also limits which
models can be analyzed.
Our attribution estimates use a permutation approximation with 200 evaluations over
20 windows, which yields a small number of permutations per item. We therefore restrict
our claims to population-level effects computed over thousands of items rather than to
per-item attributions or to small differences between models. We do not interpret
cross-model gaps in attribution mass below approximately 0.05. We note that estimation
noise attenuates our threshold-based measures: symmetric error near the $D_T = 1$
boundary moves the grounded and ungrounded fractions toward one another, so the
reported proportion of incorrect answers that are nonetheless grounded is a
conservative estimate. Quantifying estimator variance across repeated Shapley runs is left to future work.

\section{Conclusion}

TRACE evaluates whether LALM answers are grounded on task-relevant acoustic evidence and
whether that relationship survives controlled acoustic change. Pairing synthesized
scenes with event-level ground truth, programmatic answers, and matched interventions
lets the benchmark separate correctness from grounding, and invariance under
answer-preserving change from responsiveness to evidence-changing intervention.
Across five models, we show that roughly two-thirds of incorrect answers are grounded on the target event, and
target attribution mass is similar across systems whose accuracy differs by
more than twenty points. Models locate the evidence a question depends on and then
fail to interpret it, and their answers shift under perturbations that barely move
where they attend. This suggests that better event detection alone will not close the remaining gap on
ordering and counting. Separating these failure modes requires measuring acoustic evidence use
instead of only accuracy, which is what TRACE is built to support.
\bibliography{aaai2027}


\newpage
\makeatletter
\@ifundefined{ifappendixstandalone}{%
    \newif\ifappendixstandalone
    \ifx\@nodocument\relax
        \appendixstandalonefalse
    \else
        \appendixstandalonetrue
    \fi
}{}
\makeatother

\ifappendixstandalone
\documentclass[letterpaper]{article}
\usepackage[submission]{aaai2027}
\usepackage[hyphens]{url}
\usepackage{graphicx}
\urlstyle{rm}
\def\UrlFont{\rm}
\usepackage{natbib}
\usepackage{caption}
\frenchspacing
\usepackage{booktabs}
\usepackage{amsmath,amssymb}
\usepackage{xcolor}
\newcommand{\magda}[1]{{\color{red}#1}}
\newcommand{\gi}[1]{{\color{teal}#1}}
\newcommand{\yujin}[1]{{\color{violet}#1}}
\setcounter{secnumdepth}{1}
\begin{document}
\fi

\appendix
\setcounter{secnumdepth}{1}

\renewcommand{\topfraction}{0.9}
\renewcommand{\bottomfraction}{0.7}
\renewcommand{\textfraction}{0.1}
\renewcommand{\floatpagefraction}{0.75}
\setcounter{topnumber}{3}
\setcounter{bottomnumber}{2}
\setcounter{totalnumber}{4}

\section{Implementation Details}
\subsection{Seeds}
In our implementation, we only had seeds at the data generation step. We used those in our very first controlled experiments, and then removed them when generating the final benchmark, which will be shared with the acceptance of the paper. Because we release the generated benchmark itself rather than a generation seed, the scenes, questions, and annotations we evaluate on are fixed and can be reproduced exactly from the data we provide. Model inference is deterministic, so the only stochastic component left in the pipeline is the permutation sampling inside the Shapley approximation described below, which is why we restrict our claims to population-level effects rather than to per-item attributions.

\subsection{Computing Details}
For data generation, we use our own up-to-date implementation of \texttt{scaper}, which supports newer versions of Python ($\geq$ 3.12) and in which we disabled some folder-structure constraints to adapt it to our use case. Software library requirements are uploaded in the submitted code as \texttt{requirements\_hf.txt} for Qwen2.5-Omni, Qwen3-Omni, Audio Flamingo 3 and Audio Flamingo Next and \texttt{requirements\_kimi.txt} for Kimi-Audio, since they had conflicting library versions.
We ran experiments on a shared compute cluster managed via Slurm, with GPU allocation determined by resource availability at job submission time. Depending on availability, jobs were assigned one of the following GPU types: H200, L40S, H100, or A100. All computing is done with maximum memory allocation of 64GB.
We provide additional code in which we benchmark how much time models will take to run inference in different scenarios, specifically varying the number of windows and the number of \texttt{max\_evals}, which determines how many permutations we run in a given sample. We ran the benchmark on 1x A100 per model.

\subsection{Number of Runs}
As mentioned in the main paper, we use PermutationExplainer as one approximation of the Shapley Values. The number of permutations is determined by the \texttt{max\_evals} parameter.
One permutation needs $(2 \cdot \text{n\_features} + 1)$ evals to be completed. In our study, we used 20 audio windows, so $\text{n\_features} = 20$ and $\text{max\_evals} = 200$, which is between four and five complete permutations per inference ($200 / 41 \approx 4.9$).
In total, because of the computational cost, we ran only one inference per model. We parallelize our array via the \texttt{jobArray} Slurm command. This is why the analyses below are reported as population-level statistics over thousands of items, and why we do not interpret small differences between models.

\section{Benchmark Details}
\subsection{Label Distribution}
\begin{figure}[tb]
    \centering
    \includegraphics[width=\linewidth]{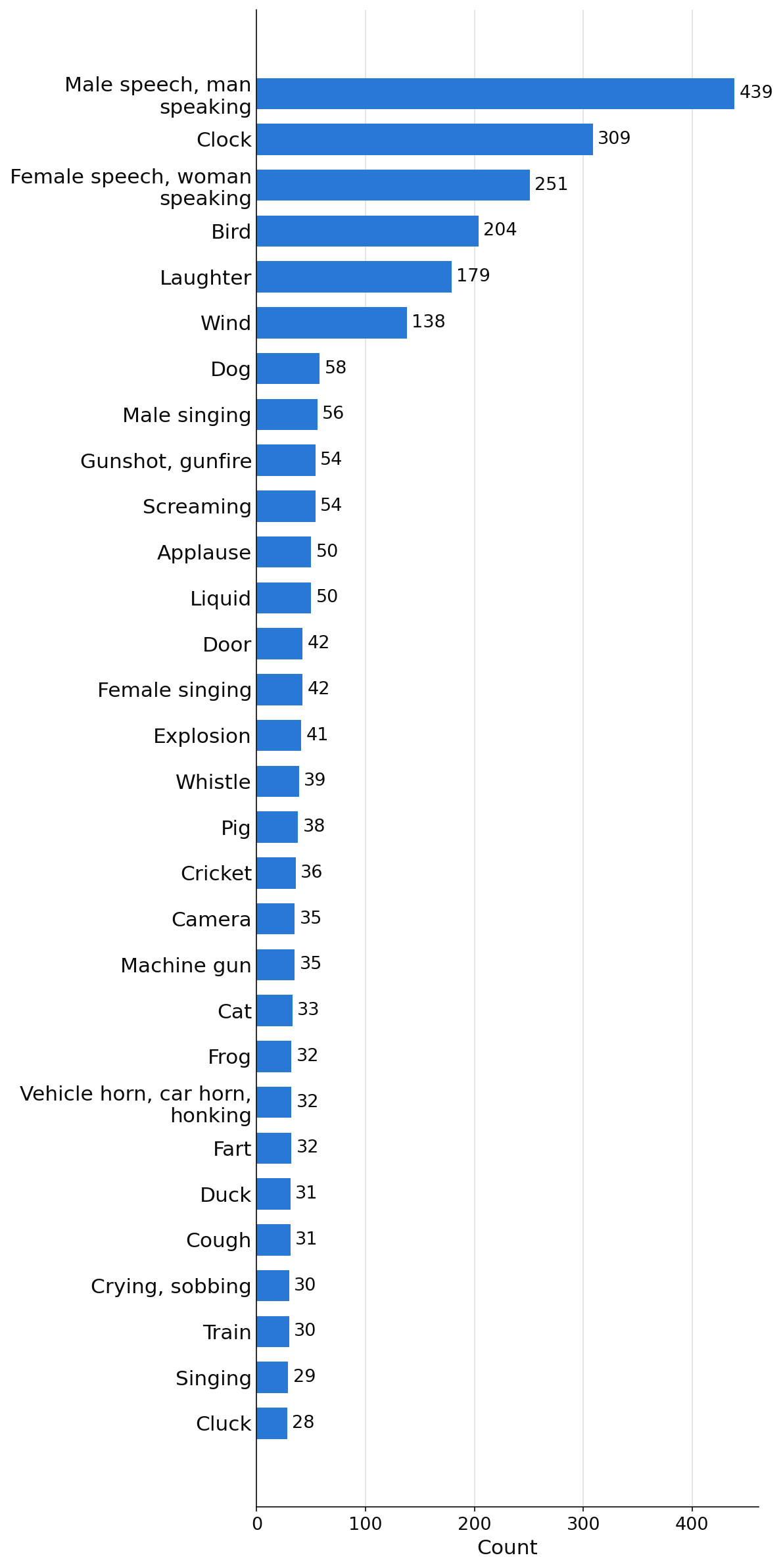}
    \caption{Distribution of labels in the dataset. The plot is showing top 30 labels from the reference tracks.}
\label{fig:label_distribution}
\end{figure}

Figure \ref{fig:label_distribution} shows the top 30 labels in the reference track. The dataset is generated with 74 unique class labels. Overall, 2,977 foreground events are generated for 1,000 reference tracks, and 8,994 foreground events are generated for 3,000 variation tracks, for a total of 11,971 foreground events in our benchmark. The 12,768 event instances reported in the main paper additionally include the background events introduced by the background variation.
This distribution is inherited from AudioSet Strong eval.

\subsection{Question Templates}
The list of questions we used for this study is categorized per question type. Existence questions have two variations: positive and negative. For positive existence questions, \texttt{\{CLASS\}} is part of the sampled classes for the scene. For negative questions, \texttt{\{CLASS\}} is not present, but it is sampled from our co-occurrence matrix, so it is an event that is likely to co-occur together with other events in the scene.

\begin{center}
\fbox{\begin{minipage}{0.95\linewidth}
\vspace{4pt}
\textbf{Existence}
\begin{itemize}
    \item ``Are there any \{CLASS\} sounds in the audio?''
\end{itemize}

\textbf{Ordering}
\begin{itemize}
    \item ``Do \{A\} sounds happen before \{B\} sounds?''
    \item ``Does \{A\} sound last?''
    \item ``Does \{A\} sound first?''
    \item ``Do \{A\} sounds and \{B\} sounds overlap in time?''
\end{itemize}

\textbf{Counting}
\begin{itemize}
    \item ``Does \{A\} occur exactly \{N\} times?''
    \item ``Does \{A\} occur more than \{N\} times?''
    \item ``Does \{A\} occur less than \{N\} times?''
\end{itemize}
\vspace{4pt}
\end{minipage}}
\end{center}

\section{Supplementary Analysis}

\subsection{Attribution in Overlapping Regions}

Target and distractor events can share analysis windows, so $A_T$, $A_D$ and $A_O$ do
not partition the attribution. To account for this, we quantify the shared portion with $A_\cap$, the
attribution mass falling in $G_T \cap G_D$, which closes the identity
$A_T^{(c)} + A_D^{(c)} - A_\cap^{(c)} + A_O^{(c)} = 1$.

$A_\cap$ becomes non-negligible only when events coincide due to a transformation: it
reaches $0.020$--$0.063$ under the overlap variation and $0.046$--$0.064$ under
repetition, against $0.004$--$0.011$ for distance and $0.000$--$0.006$ for order
inversion. Under overlap the attributed intersection is close to the intersection's own
duration share ($0.052$), showing that the shared windows are ambiguous rather than
over-attended.

Overlap is the condition in which attribution mass degrades most
consistently. Non-target foreground events attract more attribution mass than the target
in every model ($0.427$--$0.472$ vs.\ $0.323$--$0.396$). Because $A_\cap$ is credited to
both $A_T$ and $A_D$, it cancels in their difference, so the distractor advantage of
$0.067$--$0.104$ is not an artifact of shared windows.

Table~\ref{tab:background_share} reports $A_O$ for every model and variation type.
Attribution to regions outside annotated events falls below the duration baseline in
every cell, so ungrounded answers are not explained by reliance on background context.

\subsection{Sign of Attribution}

Shapley values are computed on the binary difference of positive and negative logits, i.e. $m = L_{\mathrm{yes}} - L_{\mathrm{no}}$, and are attributed to $m$ directly, without re-anchoring to the answer the model selected. A positive-valued $m$ represents the model answering \emph{yes} and a negative-valued $m$ \emph{no}; correspondingly, a window with $\phi_t > 0$ pushes the margin toward \emph{yes} and a window with $\phi_t < 0$ pushes it toward \emph{no}.
With this information, we can compare the model answers with the attribution and the ground truth, and evaluate the cases in which those disagree. We call a grounded case \emph{consistent} when the dominant sign of the target
attribution points toward the correct answer, and \emph{contradictory} when it points
away from it. A model can have high mass on the correct event while that
mass argues for the wrong response. Symmetrically, a model can answer incorrectly even though the target evidence was pointing in the right direction the whole time, which is an integration failure rather than an evidence-extraction failure.

Table \ref{tab:polarity} shows the distribution of attribution mass for the different tasks and for positive and negative questions. Overall, positive attribution dominates no matter the correct answer, and this is clearest in positive existence questions. Other types of question show a small fluctuation, with ordering questions being the only one that actually shows a negative net mass on questions that have \emph{no} as the correct answer. Ordering is also the only task in which the same event is read with the opposite sign depending on whether the model gets the question right: net attribution is $+0.166$ on correct \emph{yes} answers against $-0.026$ on incorrect ones, and $-0.051$ on correct \emph{no} answers against $+0.112$ on incorrect ones. Positive existence, in contrast, stays positive on errors and only weakens, from $+0.420$ to $+0.098$. This suggests that the models are able to identify events, but not necessarily use them for interpretation.

Counting is the exception. When the correct answer is \emph{no}, net attribution has a small, but positive value on the wrong sign ($+0.019$), and only two of the five models show the expected sign value. A count is a property of a set of instances rather than of any single one of them, so no individual interval should be expected to push the margin in a consistent direction, and we do not claim a coherent sign structure for counting.

\begin{table}[tb]
\centering
\small
\setlength{\tabcolsep}{3.5pt}
\begin{tabular}{llrrrrrr}
\toprule
& & \multicolumn{3}{c}{\textbf{Correct}} & \multicolumn{3}{c}{\textbf{Incorrect}} \\
\cmidrule(lr){3-5}\cmidrule(lr){6-8}
\textbf{Task} & \textbf{Ans.} & pos & neg & net & pos & neg & net \\
\midrule
existence & yes & 0.445 & 0.025 & +0.420 & 0.191 & 0.093 & +0.098 \\
order & yes & 0.223 & 0.057 & +0.166 & 0.105 & 0.131 & $-$0.026 \\
order & no & 0.089 & 0.140 & $-$0.051 & 0.177 & 0.066 & +0.112 \\
counting & yes & 0.161 & 0.110 & +0.051 & 0.087 & 0.213 & $-$0.126\\
counting & no & 0.139 & 0.120 & +0.019 & 0.264 & 0.067 & +0.197 \\
\bottomrule
\end{tabular}
\caption{Positive and negative attribution mass on the target event, pooled over models
and split by the correct answer and by whether the model answered correctly.
net~$=$~pos~$-$~neg. The expected pattern is net~$>0$ when the correct answer is
\emph{yes} and net~$<0$ when it is \emph{no}. Negative-existence questions are omitted because they have no target region.}
\label{tab:polarity}
\end{table}

\begin{table}[tb]
\centering
\footnotesize
\setlength{\tabcolsep}{3pt}
\begin{tabular}{lrrrrrr}
\toprule
\textbf{Variation} & \textbf{Dur.} & \textbf{AF3} & \textbf{AF-Next} & \textbf{Kimi} & \textbf{Qwen2.5} & \textbf{Qwen3} \\
\midrule
Removal        & 0.641 & 0.468 & 0.434 & 0.545 & 0.443 & 0.432 \\
Overlap        & 0.554 & 0.227 & 0.203 & 0.270 & 0.197 & 0.202 \\
Distance       & 0.477 & 0.311 & 0.296 & 0.348 & 0.290 & 0.268 \\
Inv. order     & 0.477 & 0.318 & 0.306 & 0.343 & 0.294 & 0.276 \\
Background     & 0.456 & 0.363 & 0.357 & 0.403 & 0.369 & 0.368 \\
SNR            & 0.443 & 0.327 & 0.321 & 0.361 & 0.311 & 0.302 \\
Repetition     & 0.343 & 0.184 & 0.160 & 0.173 & 0.146 & 0.150 \\
\bottomrule
\end{tabular}
\caption{Attribution mass assigned to regions outside annotated events,
compared with those regions' share of clip duration (Dur.). The duration share is the
fraction of the clip lying outside the union of target and distractor events,
so it accounts for the shared portion $A_\cap$. Every model in every
condition falls below the duration baseline: models under-weight background
rather than relying on it.}
\label{tab:background_share}
\end{table}

\subsection{Grounding and Cross-Model Accuracy}

\begin{figure}[!htb]
    \centering
    \includegraphics[width=0.85\linewidth]{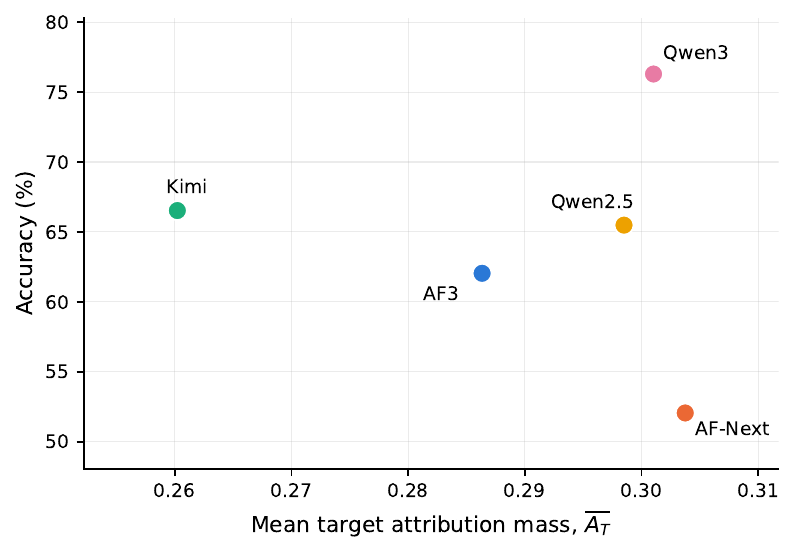}
    \caption{Accuracy versus mean target attribution mass. Points show model means.
    Higher grounding does not determine higher performance across models.}
    \label{fig:scatter}
\end{figure}

Target attribution mass is nearly constant across models ($0.259$--$0.306$) while
accuracy varies by more than twenty points ($55.0$--$78.4\%$). Attribution mass does not
explain the accuracy differences between the evaluated models: all five allocate a similar share of attribution to the target region and differ in
what they do with it. Figure~\ref{fig:scatter} makes the dissociation explicit.
AF-Next has the lowest accuracy but among the highest $A_T$; Kimi is more accurate
than Qwen2.5 with lower $A_T$ and lower target density; Qwen3 improves accuracy
without exceeding AF-Next or Qwen2.5 in $A_T$.

This dissociation is between models, but not within them.
We tested whether target attribution predicts answer
correctness with a binomial mixed-effects logistic regression, \texttt{correct
$\sim$ $A_T$}, with a random intercept per reference scene to account for the fact
that a scene and its variations are not independent observations. We fit one model
per system, since attribution magnitude and distribution differ
substantially across systems. The fixed effect of $A_T$ was positive for every system, with
odds ratios from $1.86$ ($95\%$ CI $[1.57, 2.21]$, Qwen3) to $18.63$
($[15.99, 21.70]$, AF-Next), so higher target attribution is associated with
higher odds of a correct answer at the item level. These odds ratios are expressed per unit of $A_T$, which is bounded in $[0,1]$ and whose realized spread is roughly $0.2$, so they describe a change larger than any item actually undergoes; over a more representative change of $0.1$ in $A_T$ the odds ratio is $1.06$--$1.34$. The effect is therefore positive but modest.

Together with the population-level result that all models place significantly more
mass on target regions than their duration predicts, this indicates
above-chance sensitivity to the relevant acoustic evidence, but that alone is not enough
to separate systems.

\section{Visualizing Individual Examples}
Besides the metrics we compute aggregating the Shapley Values, we also performed qualitative analysis and inspected individual examples. To do so, we implemented a small explorer application that allows us to compare different plots, listen to examples and filter by our metrics.
Below we show two kinds of comparison. We can visualize the attribution values via heatmaps, together with the event roll, which lets us read whether the mass sits on the target event, on a distractor, or outside both, and how that allocation moves across the variations of the same scene.

\begin{figure}[htb]
    \centering
    \includegraphics[width=\linewidth]{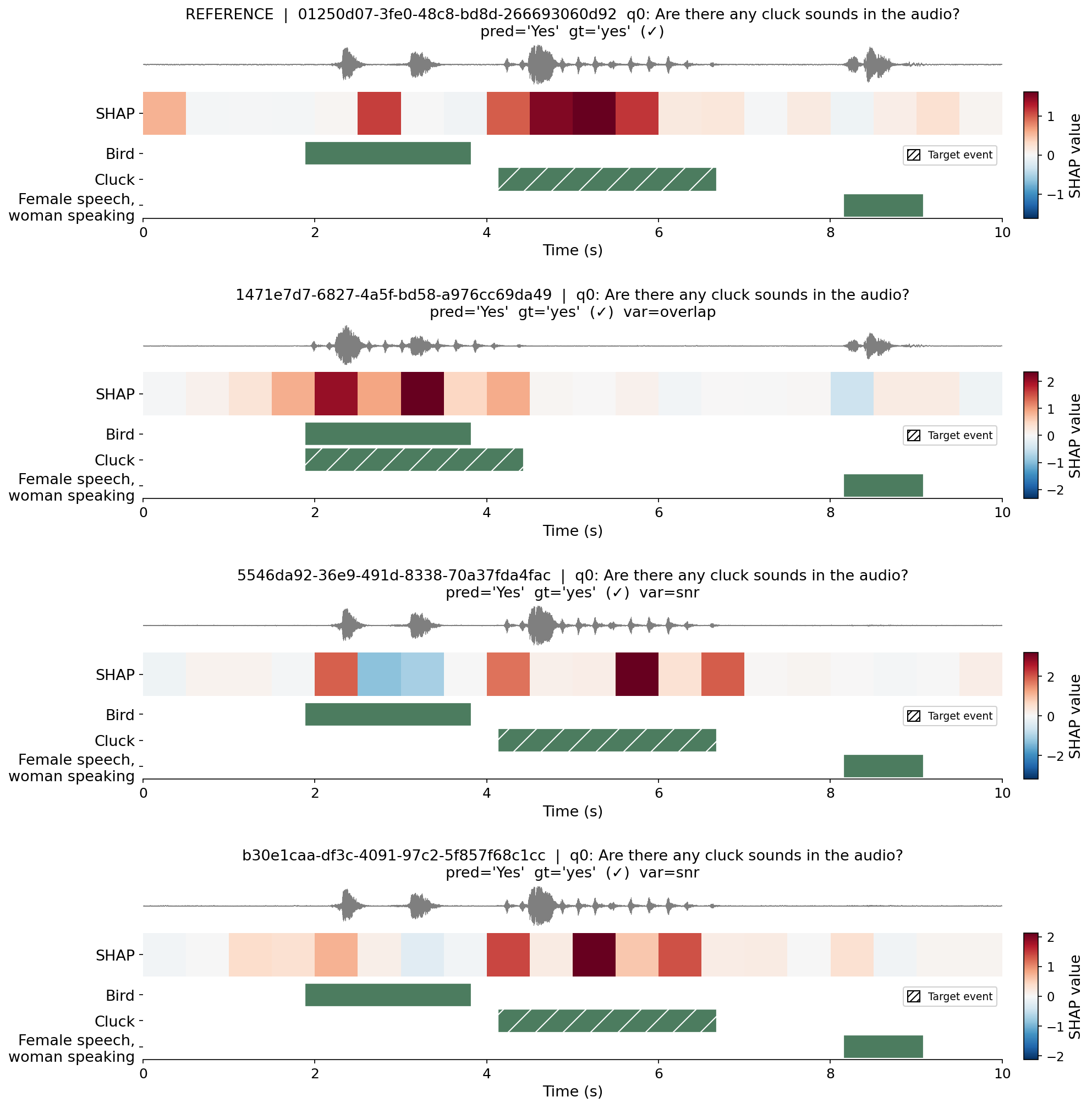}
    \caption{Qualitative analysis example for Audio Flamingo 3. In the first row, we have the reference scene and what is the question the model receives, as well as the model prediction and answer. Each plot has the Shapley values for each audio token and the sound event roll with target and distractor events. Each different plot is showcasing how the attribution changes for a given variation.}
\label{fig:af3_q00}
\end{figure}

Figure \ref{fig:af3_q00} shows the reference scene for Audio Flamingo 3 in the first row, followed by each of the variations we apply to it. This type of plot allows us to see how much the variations affect the model and how the attribution shifts.

\begin{figure*}[tb]
    \centering
    \includegraphics[width=\linewidth]{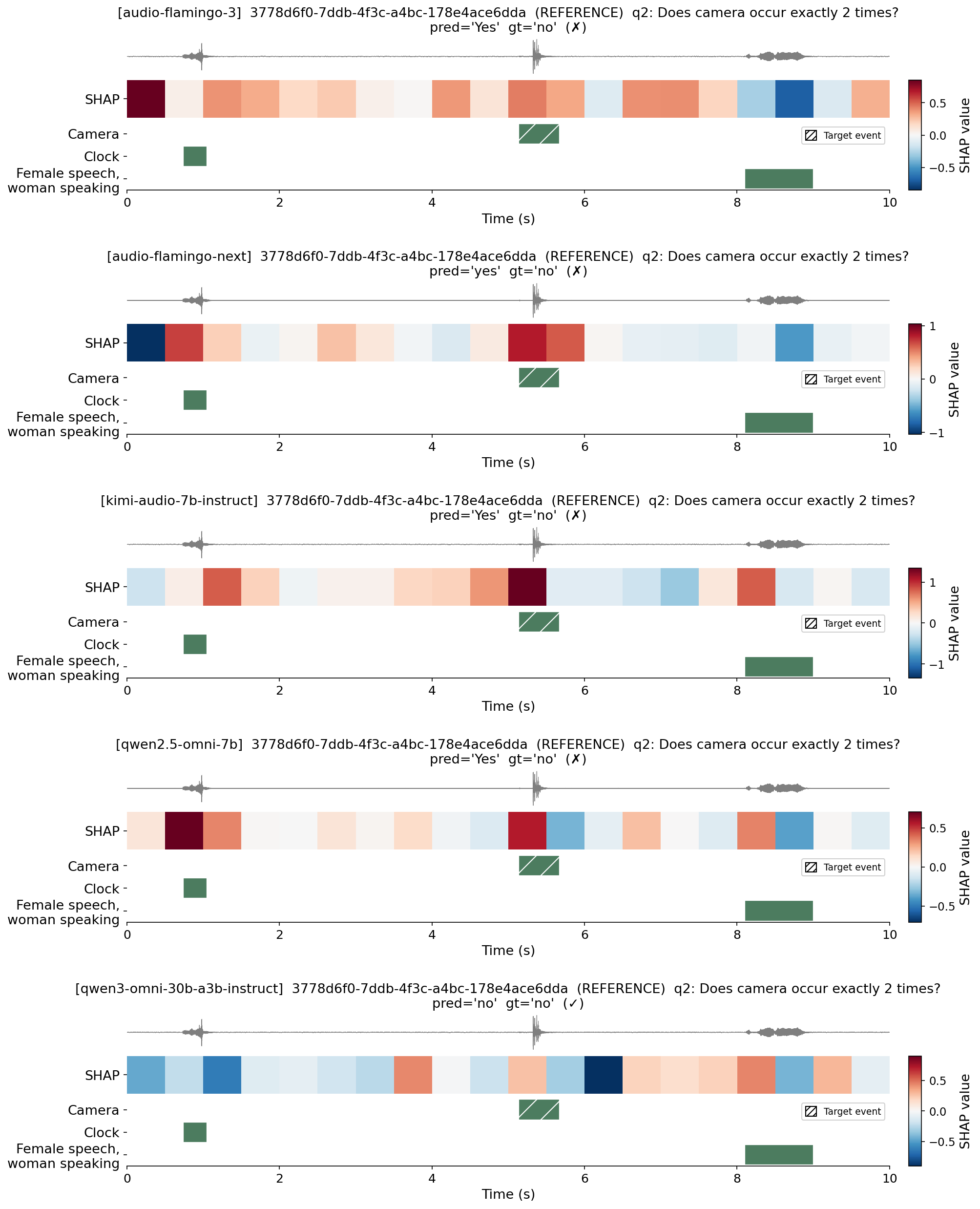}
    \caption{Model comparison for the same reference question. Each heatmap represents one model and its attributions.}
\label{fig:q02_reference}
\end{figure*}

\begin{figure*}[tb]
    \centering
    \includegraphics[width=\linewidth]{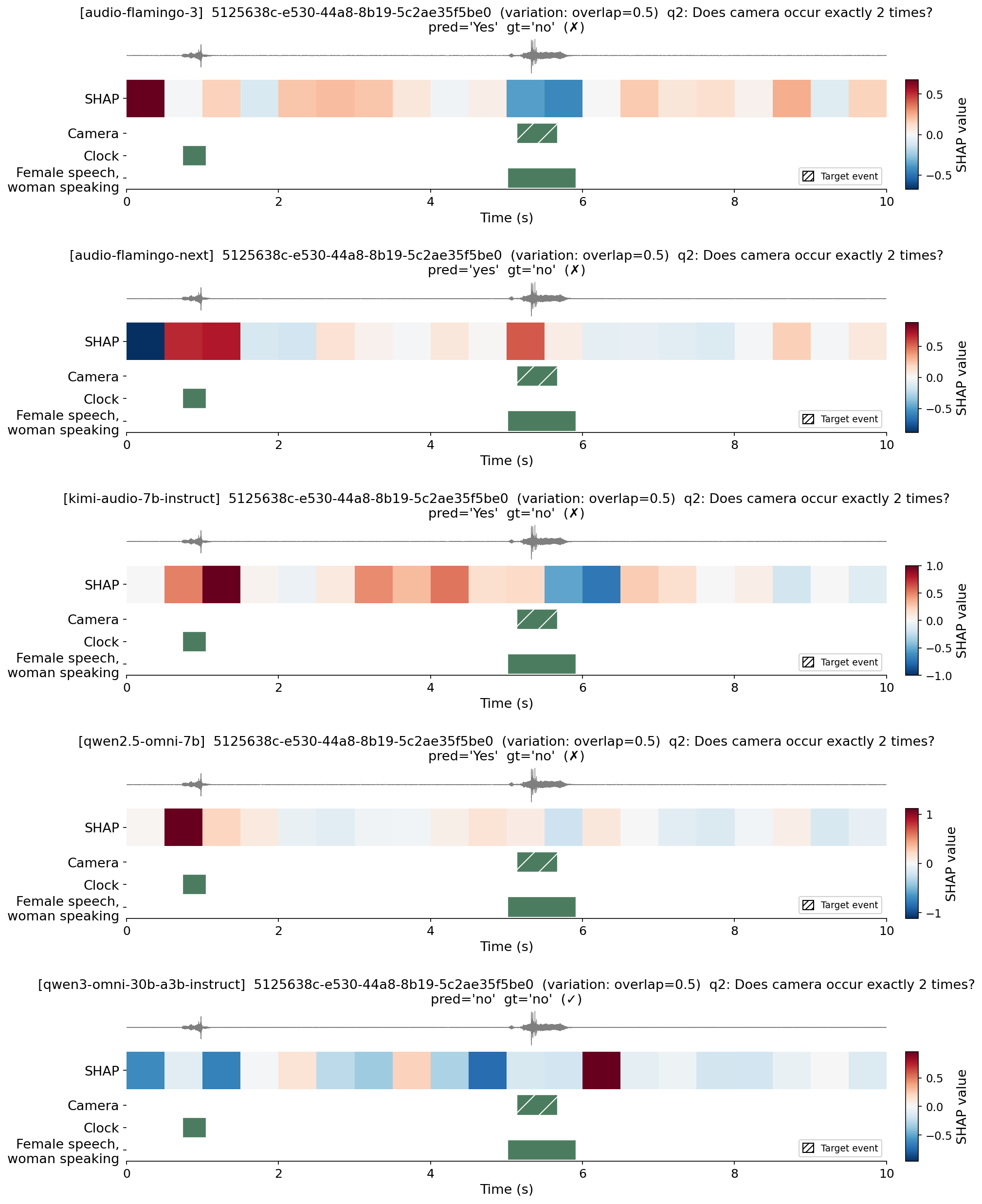}
    \caption{Model comparison for the same question as Figure \ref{fig:q02_reference}, but applying the overlap variation.}
\label{fig:q02_overlap}
\end{figure*}

Figures \ref{fig:q02_reference} and \ref{fig:q02_overlap} are examples of comparing the attributions between models.
Because of space constraints, we will share more plots in a demo page. Figures \ref{fig:q02_reference} and \ref{fig:q02_overlap} have sounds that are different in category (clock and camera), but similar in sound (a click). As a result, we can see how some of the attributions assign a high value to the ``clock'' class, especially when making the wrong prediction for the question.

\ifappendixstandalone
\end{document}
\fi 

\end{document}